\documentclass[twocolumn,twocolappendix]{aastex631}
\usepackage{multirow}
\usepackage{amsmath}
\usepackage{booktabs}
\usepackage{floatrow}

\usepackage{xcolor}

\newcommand\lsim{\mathrel{\rlap{\lower4pt\hbox{\hskip1pt$\sim$}}\raise1pt\hbox{$<$}}}
\newcommand\gsim{\mathrel{\rlap{\lower4pt\hbox{\hskip1pt$\sim$}}\raise1pt\hbox{$>$}}}

\newcommand{\jwst}[0]{\textit{JWST }}
\newcommand{\jwstns}[0]{\textit{JWST}}
\newcommand{\lcdm}[0]{\text{$\Lambda$CDM }}

\accepted{ApJL}
\shorttitle{Stream velocity with JWST}
\shortauthors{Williams et al.}
\begin{document}

\title{The Supersonic Project: Lighting up the faint end of the JWST UV luminosity function}


\correspondingauthor{Claire E. Williams}
\email{clairewilliams@astro.ucla.edu}

\author[0000-0003-2369-2911]{Claire E. Williams}
\affil{Department of Physics and Astronomy, UCLA, Los Angeles, CA 90095}
\affil{Mani L. Bhaumik Institute for Theoretical Physics, Department of Physics and Astronomy, UCLA, Los Angeles, CA 90095, USA\\}

\author[0000-0002-4227-7919]{William Lake}
\affil{Department of Physics and Astronomy, UCLA, Los Angeles, CA 90095}
\affil{Mani L. Bhaumik Institute for Theoretical Physics, Department of Physics and Astronomy, UCLA, Los Angeles, CA 90095, USA\\}

\author[0000-0002-9802-9279]{Smadar Naoz}
%\author{Smadar Naoz}
%\affil{Department of Physics and Astronomy, University of California, Los Angeles, CA 90095\\}
\affil{Department of Physics and Astronomy, UCLA, Los Angeles, CA 90095}
\affil{Mani L. Bhaumik Institute for Theoretical Physics, Department of Physics and Astronomy, UCLA, Los Angeles, CA 90095, USA\\}

\author[0000-0001-5817-5944]{Blakesley Burkhart}
%\author{Blakesley Burkhart}
\affiliation{Department of Physics and Astronomy, Rutgers, The State University of New Jersey, 136 Frelinghuysen Rd, Piscataway, NJ 08854, USA \\}
\affiliation{Center for Computational Astrophysics, Flatiron Institute, 162 Fifth Avenue, New York, NY 10010, USA \\}

\author[0000-0002-8460-0390]{Tommaso Treu}
%\author{Smadar Naoz}
%\affil{Department of Physics and Astronomy, University of California, Los Angeles, CA 90095\\}
\affil{Department of Physics and Astronomy, UCLA, Los Angeles, CA 90095}

\author[0000-0003-3816-7028]{Federico Marinacci}
%\author{Federico Marinacci}
\affiliation{Department of Physics \& Astronomy ``Augusto Righi", University of Bologna, via Gobetti 93/2, 40129 Bologna, Italy\\}

\author[0000-0002-0984-7713]{Yurina Nakazato}
\affiliation{Department of Physics, The University of Tokyo, 7-3-1 Hongo, Bunkyo, Tokyo 113-0033, Japan}

\author[0000-0001-8593-7692]{Mark Vogelsberger}
%\author{Mark Vogelsberger}
\affil{Department of Physics and Kavli Institute for Astrophysics and Space Research, Massachusetts Institute of Technology, Cambridge, MA 02139, USA\\}

\author[0000-0001-7925-238X]{Naoki Yoshida}
\affiliation{Department of Physics, The University of Tokyo, 7-3-1 Hongo, Bunkyo, Tokyo 113-0033, Japan}
\affiliation{Kavli Institute for the Physics and Mathematics of the Universe (WPI), UT Institute for Advanced Study, The University of Tokyo, Kashiwa, Chiba 277-8583, Japan}
\affiliation{Research Center for the Early Universe, School of Science, The University of Tokyo, 7-3-1 Hongo, Bunkyo, Tokyo 113-0033, Japan}

\author[0000-0001-6246-2866]{Gen Chiaki}
\affiliation{Astronomical Institute, Tohoku University, 6-3, Aramaki, Aoba-ku, Sendai, Miyagi 980-8578, Japan}

\author[0000-0003-4962-5768]{Yeou S. Chiou}
%\author{Yeou S. Chiou}
\affil{Department of Physics and Astronomy, UCLA, Los Angeles, CA 90095}
\affil{Mani L. Bhaumik Institute for Theoretical Physics, Department of Physics and Astronomy, UCLA, Los Angeles, CA 90095, USA\\}

\author[0000-0002-8859-7790]{Avi Chen}
\affiliation{Department of Physics and Astronomy, Rutgers, The State University of New Jersey, 136 Frelinghuysen Rd, Piscataway, NJ 08854, USA \\}


\begin{abstract}
The {\it James Webb Space Telescope} ({\it JWST}) is capable of probing extremely early eras of our Universe when the supersonic relative motions between dark matter and baryonic overdensities 
modulate structure formation  ($z\gsim 10$). 
We study low-mass galaxy formation including this ``stream velocity" using high resolution {\tt AREPO} hydrodynamics simulations, and present 
theoretical predictions of the UV luminosity function (UVLF) and galaxy stellar mass function (GSMF) down to extremely faint and low mass galaxies ($M_{UV}\gsim -15$, $10^4M_\odot \leq M_*\leq 10^8 M_\odot)$.
We show that, although the stream velocity suppresses early star formation overall, it induces a short period of rapid star formation in some larger dwarfs, leading to an enhancement in the faint-end of the UVLF at $z=12$.
We demonstrate that \jwst observations are close to this enhanced regime, and propose that the UVLF may constitute an important probe of the stream velocity at high redshift for \jwst and future observatories.

\end{abstract}


\section{Introduction}
The \textit{James Webb Space Telescope} ({\it JWST}) has opened a new window on the first galaxies in the Universe, allowing for new insights regarding their properties and formation. 
These galaxies at ``Cosmic Dawn" host the yet undetected first generation of stars (Population III), reionize the Universe, and enrich their surroundings with the first metals.
The \lcdm (cold dark matter + cosmological constant) cosmological model predicts the formation of these early sources within the first few hundred million years after the Big Bang \citep[e.g.,][]{Vogelsberger+20}{}{}. 
Observations of faint galaxies at high redshift represent an opportunity to test galaxy formation models and the \lcdm paradigm itself. 

Since \jwst began science operations in 2022, NIRCam datasets have provided a number of intriguing galaxy candidates at $z>9$ \citep[e.g.,][]{Naidu+22candidates,Castellano+22glass,Castellano+23glass2,Adams+23,Atek+23candidates,Finkelstein+22z12candidate,Yan+23,Donnan+23b,Donnan+23uvlf,Morishita+23,Bradley+22,Bouwens+23uvlf,PerezGonzalez+23}.
Furthermore, \jwstns's spectrographic instruments allow for precision redshift determination, and multiple galaxies with $z>9$ have been confirmed, with gravitationally lensed systems providing an additional magnified glimpse at the early Universe \citep[e.g.,][]{Roberts-Borsani+22,Williams+22,Atek+23candidates,ArribalHaro+23}. 
Moreover, robust measurements of the 
UV luminosity function
(UVLF) are now possible down to very faint magnitudes \citep[e.g.,][]{Naidu+22candidates,Leung+23uvlf,Castellano+23glass2,Finkelstein+23ceersI,Donnan+23uvlf,Harikane+23spec,Harikane+23uvlf}{}{} at redshifts $z\sim12$ and beyond.
These observations open a window into the physics of galaxy formation at increasingly small scales and high redshifts.

Low-mass (baryonic mass ($M_b$) $\lsim 10^{9} M_\odot$) galaxies  represent a sensitive test of galaxy formation and cosmological models due to their small gravitational potentials, which are sensitive to the precise nature of dark matter (DM) and baryonic physics \citep[e.g.,][]{BullockBK+17,Sales+22}.
In order to compare the wealth of \jwst data to cosmological models, a precise description of the baryonic physics that dictates the formation of these dwarf galaxies is necessary \citep[e.g.,][]{Vogelsberger+20JWST,Shen+20JWST,Shen+23UV}{}{}.
Within the standard \lcdm model, these galaxies originate from small scale
($k\gsim 20$ Mpc$^{-1}$)
 baryonic density fluctuations, whose growth is dictated by the overwhelming gravitational potential of DM overdensities. The latter is orders of magnitude larger than baryonic overdensities 
 by the time of recombination \citep[e.g.,][]{NB05,LoebFurlanetto+13}. 

\citet{Tes+10a} showed that a second-order term in standard linear perturbation theory, the relative velocity between dark matter and baryons ($v_{\rm bc}$), is non-negligible at these scales, with an rms value ($\sigma_{\rm vbc}$) at $z=1100$ of $30$~km~s$^{-1}$, five times the speed of sound at  that time. 
This stream velocity was shown to delay the formation of Pop III stars, suppress halo abundance, and raise the minimum halo mass that can retain baryons and form stars \citep[e.g.,][]{Stacy+10,Greif+11,Schauer+17a,Schauer+21, Schauer+23,Naoz+12, BD, Asaba+16,Nebrin+23,Hegde+23,Conaboy+23}. Notably, the stream velocity was recently estimated locally to be $v_{\rm bc}=1.75^{+0.13}_{-0.28}\sigma_{\rm vbc}$ \citep{Uysal+22}.

This supersonic velocity generates a phase shift between the baryonic and dark matter perturbations that introduces a physical offset between the center of mass of collapsed baryon structures and their parent DM halo \citep[e.g.,][]{Naoz+14,Popa+15,Chiou+18,Chiou+19,Chiou+21,Lake+21,Nakazato+22,Lake+22,Williams+23,Lake+23b}. 
The physical offset between DM and baryonic components of \lcdm overdensities and other effects of the stream velocity have important implications for the formation of the first stars and galaxies, with  potential consequences for \jwst observations. 
 In some cases for very low mass halos ($M_b\lsim10^5 M_\odot$), the physical offset is so large that baryons collapse outside of their parent DM halo, creating DM-free Supersonically-Induced Gas Objects (SIGOs) which may be the progenitors of some early globular clusters \citep[e.g.,][]{Popa+15,Chiou+18,Chiou+19,Chiou+21,Lake+21,Nakazato+22,Lake+22}. 
Future \jwst observations may reveal a population of star clusters at high galactocentric distances descended from SIGOs at high redshift \citep[][]{Lake+23b}{}{}.
For halos with $M_b\lsim10^{9} M_\odot$, the physical offset results in the formation of diffuse structures called Dark Matter + Gas Halos Offset by Streaming (DM GHOSts) \citep[e.g.,][]{Williams+23}. Unlike typical dwarf galaxies, these structures are highly elongated, more rotationally supported, and gas-deficient, as the stream velocity advects a portion of the gas component out of the halo \citep[e.g.,][]{Williams+23,Hirano+23}.

The delay of star formation and the diffuse nature of the structures that form in regions of streaming have implications for the populations of dwarf galaxies that exist in the high redshift Universe. 
Here, we suggest that the stream velocity may affect the faint end of the UVLF by suppressing star formation in the smallest galaxies and boosting the star formation rate (SFR) of larger dwarfs for a short period around $z\sim 12$.  
We estimate the galaxy stellar mass function (GSMF), showing that the stream velocity suppresses the low mass end. 
We suggest that \jwst and future extremely large telescopes (ELTs) have the opportunity to observe galaxies at extremely high redshift in regions of supersonic streaming. 
These results imply that 
the stream velocity will enhance the scatter in the UVLF at faint $M_{UV}$.

The structure of this paper is as follows: in \S~\ref{sec:Methods} we describe the {\tt AREPO} simulations used and their initial conditions including the stream velocity. 
In \S~\ref{sec:Results}, we describe the galaxy stellar mass function (\S~\ref{sec:GSMF}) and the UV luminosity function (\S~\ref{sec:uvluminosity})  at high redshifts in regions of the universe with and without streaming. 
Finally, in \S~\ref{sec:Discussion} we interpret these results and the outlook for \jwst observations. 

In this study we assume a \lcdm cosmology, with $\Omega_{\rm \Lambda} = 0.73$, $\Omega_{\rm m} = 0.27$, $\Omega_{\rm b} = 0.044$, $\sigma_8  = 1.7$, and $h = 0.71$.
Magnitudes are calculated using the AB magnitude system; $m=-2.5\log_{10}(f_\nu /\text{ nJy}) + 31.40$ \citep[][]{Oke-74}. 

\section{Methods}
\label{sec:Methods}
\subsection{Cosmological Simulation Description}
We perform a suite of numerical simulations in {\tt AREPO} \citep[][]{Springel2010a}, evolving a box (2.5 h$^{-1}$ Mpc)$^3$ from $z=200$ to $z=12$.
To generate initial conditions, we use a modified CMBFAST code \citep{1996ApJ...469..437S}. As in \citet{Popa+15},  we include the first-order correction of scale-dependent temperature fluctuations on the initial conditions and their transfer functions as in \citet[][]{NB05,Naoz+13}. 
This allows for accurate gas fractions in halos at high redshift \citep[e.g.,][]{NBM,Naoz+10,Naoz+12}.
The box contains 768$^3$ DM particles ($m_{\rm DM}=775 h ^{-1} M_\odot$) and 768$^3$ Voronoi mesh gas cells, giving a gas cell resolution of $m_{\rm gas}= 200\,M_\odot$. See \citet[][]{Lake+23b}{}{} for the simulation suite details.

On scales smaller than a few Mpc, the relative velocity is coherent and can be modeled as a bulk stream motion \citep[e.g.,][]{Tes+10a,Popa+15} with an rms value of $\sigma_{\rm bc}$.
We initialize our simulations at $z=200$, when a 2$\sigma_{\rm bc}$ fluctuation in the stream velocity corresponds to $11.8$~km~s $^{-1}$ \citep[e.g.,][]{Popa+15}{}{}.
The stream velocity is thus implemented as a uniform boost of 11.8 ( km s$^{-1} ) \hat{x}$ to all baryon particles. 
In order to compare the properties of early galaxies in regions with and without a highly supersonic stream velocity, we perform two simulation runs, one with a $0\sigma_{\rm bc}$ fluctuation (i.e., no stream velocity) and one with a $2\sigma_{\rm bc}$ fluctuation (a value similar to the observed local one; \citealt{Uysal+22}).

In order to increase the statistical power of our simulation without affecting the underlying cosmology, we choose $\sigma_8=1.7$ (these are the same simulations as in \citealt{Lake+23b}). 
Thus, our box represents a strong density peak where structure in the Universe forms early, such as in the Virgo cluster \citep[e.g.,][]{NB07}. 
These results and their robust statistics can thus be scaled to other regions of the Universe accordingly \citep[e.g.,][]{Naoz+13,Park+20} (see App.~\ref{ap:Clustering}).
% App. \ref{ap:sig8}). 

\subsection{Subgrid Physics - Cooling and Star Formation}

An accurate prescription of molecular cooling in the pristine gas of the early Universe is necessary to model the formation and collapse of gas-rich structures at high redshift \citep[e.g.,][]{Schauer+21,Nakazato+22,Lake+22,Williams+23}. 
Following \citet{Nakazato+22, Lake+22, Williams+23}, we include molecular cooling in both runs through the GRACKLE chemistry and cooling library \citep[][]{Smith+17,Chiaki+19}. 
This prescription explicitly accounts for nonequilibrium chemical reactions and radiative cooling, including 15 primordial species (e$^-$, H, H$^+$, He, He$^+$, He$^{++}$, H$^-$, H$_2$, H$_2^+$, D, D$^+$, HD, HeH$^+$, D$^-$, and HD$^+$). 

Once gas condenses sufficiently, the first generation of stars forms. 
As described in \citet{Lake+23b}, when a gas cell exceeds the Jeans mass, it collapses to form a star particle on the free-fall timescale. 
The star formation rate (SFR) for a given object is computed as the difference in its stellar mass between two snapshots, excluding any star particles that were part of a cluster that merged into the halo. 
We do not include any stellar feedback effects, although feedback processes are vital to understand  the efficiency of star formation over cosmic time to the present day. 
The resulting estimates of the UVLF can thus be considered an upper limit, given that feedback effects will suppress the efficiency of star formation.

\subsection{Identification of structures}
Using a friends-of-friends (FOF) algorithm, we search for two object classes. 
First, we identify DM-primary/gas-secondary objects, corresponding to DM GHOSts. In this case, the FOF algorithm is run on DM particles first, with gas cells and star particles linked at a secondary stage. 
We also identify gas-primary objects by running the FOF algorithm on gas cells only.
This allows us to find gas clumps without an associated DM halo \citep[(SIGOs), see e.g.,][]{Chiou+18,Chiou+19,Chiou+21,Lake+21,Nakazato+22,Lake+22,Lake+23b}{}{}.
In order to avoid non-physical numerical effects, we require DM-primary objects to contain at least 300 particles and gas-primary objects to have at least 100 cells \citep[e.g.,][]{Naoz+10}.

\begin{figure*}[t!]

\centering
   
\includegraphics[width=0.98\linewidth]{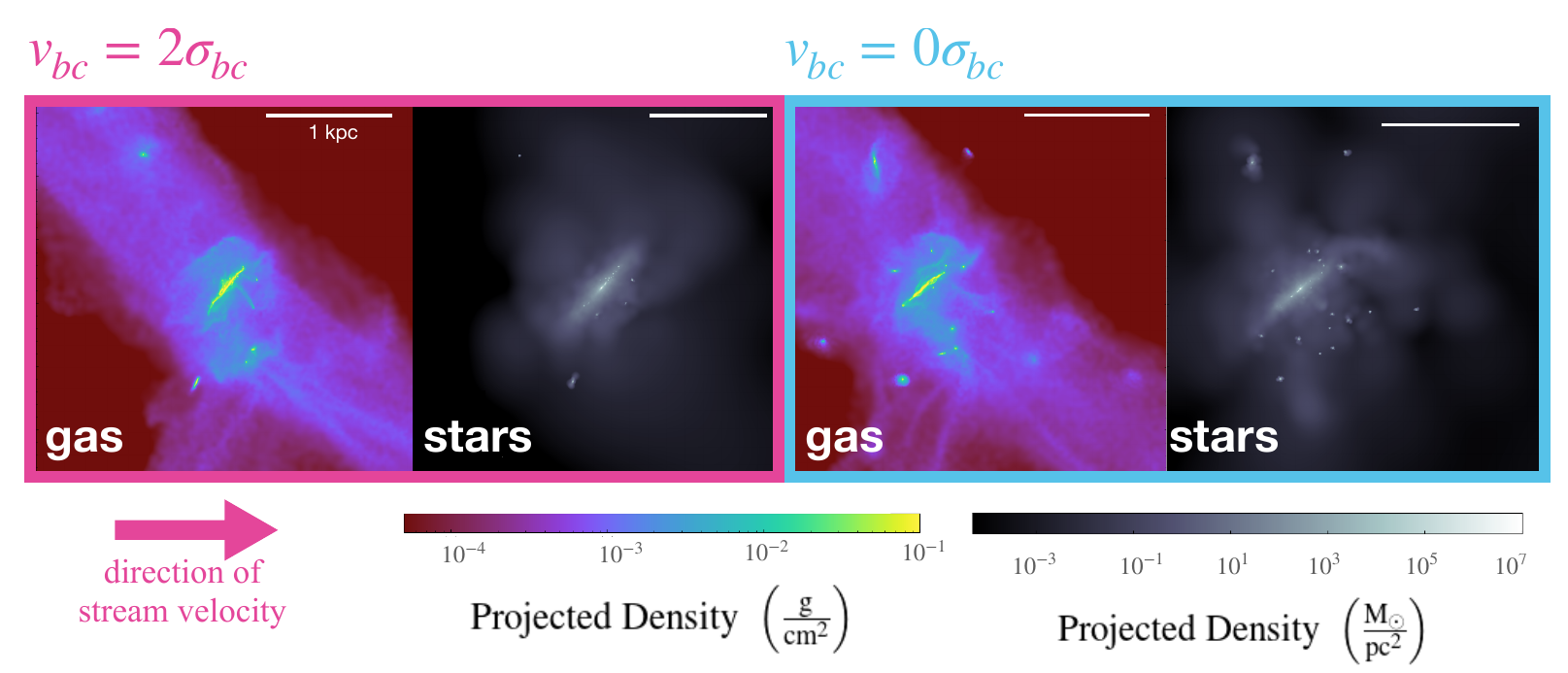}    
    \caption{Comparison of the projected density of gas and stars in the region of a highly star forming object in our simulations at $z=12$ between the $2\sigma_{\rm bc}$ run (left) and the $0\sigma_{\rm bc}$ (right) run.
    A bar is shown in each panel denoting a distance of 1 kpc.
    Both the left and right boxes are centered on a galaxy in the $2\sigma_{\rm bc}$--and a physical offset of the location of the galaxy is noticeable as the $0\sigma_{\rm bc}$ galaxy appears to the left of the box.
    This is due to the stream velocity, which points in the $+\hat{x}$ direction (to the right). 
    The galaxy at the center of the left two panels formed stars rapidly between $z=15$ and $z=12$, with a star formation rate of $\dot{M_*}=0.176\,M_\odot$ yr$^{-1}$, the fifth highest in our $2\sigma_{\rm bc}$ run. 
    It has $M_*=10^{7.51} M_\odot$ at $z=12$. The stellar mass history of this object in the interval between   $z=15$ and $z=12$ is shown in Fig.~\ref{Fig:Mstarobj} in App.~\ref{ap.stellarmass}.  
    }\label{Fig:box} 
\end{figure*}

\subsection{The UV luminosity}

The rest-frame UV $1500$\AA\text{  } luminosity of high redshift galaxies observed in the infrared by \jwst ~traces the star formation rate (SFR) assuming that they are dominated by young, massive stars with short lifetimes \citep[e.g.,][]{Kennicutt+98,Stark+09,Kennicutt+12}{}{}. 
We calculate the SFR of each object in the simulation by identifying objects in snapshots at $z=13$ and $z=12$, and calculating the difference in stellar mass between the two snapshots.
Given the SFR of galaxies in our simulations at $z=12$, we then estimate their rest-frame UV continuum luminosity ($L_{UV,1500}$): 
\begin{equation}
    L_{UV, 1500}=\frac{\dot{M}_*}{{\cal K}_{UV,1500}} \ ,
    \label{eq:Luv}
\end{equation}
where $\dot{M}_*$ is the star formation rate in units of $M_\odot$~yr$^{-1}$. 
 ${\cal K}_{UV,1500}$ is a fiducial constant evaluated for continuous-mode star formation with a Salpeter IMF. We adopt  ${\cal K}_{UV,1500}=1.15 \times 10^{-28}$ ( $M_\odot$ yr$^{-1}$ / (ergs s$^{-1}$ Hz$^{-1}$) \citep[following e.g.,][]{Madau+14,Sun+16}.

This seems to be a reasonable assumption given that we trace the star formation on a timescale of $\sim30$ Myr, which is comparable to the lifetime of the most massive stars. 
${\cal K}_{UV,1500}$ depends on the IMF of the stellar population, which remains uncertain at high redshift.
Proposed high-redshift and Pop III IMFs may shift the magnitude at which galaxies of a given mass enter the observable magnitudes of 
\jwstns.
In particular, several studies propose that a top heavy IMF may be present at high redshift, Pop III environments due to their low metallicity and/or high CMB temperature \citep[e.g.,][]{Omukai+05,Hirano+14,Hirano+15,Chon+22}{}{}. 
Such an IMF would lead to increased UV luminosity from nebular emission, increasing our $L_{UV, 1500}$ for a given $\dot{M}_*$.
Thus, we adopt the standard value of ${\cal K}_{UV,1500}$ both as a conservative estimate of detectability and for ease of comparison with other works.
This work assumes that the stream velocity does not affect the high-redshift IMF, and further work is needed to ascertain whether such a dependence exists. 
Under this assumption, the shape of the UVLF should be consistent between the stream velocity and non-stream velocity scenarios, although the overall height of both curves may vary given assumptions about ${\cal K}_{UV, 1500}$.

We must also correct our number counts for the fact that we have artificially increased $\sigma_8$ to $1.7$ in order to gain more complete statistics when calculating number counts per volume in this study.
We calculate a conversion factor from $\sigma_8=1.7$ to $\sigma_8=0.8$ as a function of mass using the \cite{ST+02} halo mass function (see {App.~\ref{ap:Clustering}}).
% \ref{ap:sig8}). 
In each magnitude bin, we divide the number counts by the correction factor corresponding to the average mass in each bin. 
This introduces an additional uncertainty in the UVLF because of the range of halo masses represented at each magnitude.

\section{Results}
\label{sec:Results}

\subsection{The Effect of Streaming on Star-Forming Regions }

As expected, the small-scale structure is suppressed in the presence of stream velocity \citep[e.g.,][]{Tes+10a, Fialkov+11,BD,Maio+11, OLMc12,Naoz+11a,Tes+10b,TanakaLi+13,Tanaka+14}{}{}.  Figure \ref{Fig:box} shows an example of the projected density of a highly star-forming region with (left panels) and without (right panels) streaming. Both cases show the gas and star components separately (see labels).
As depicted,  small clumps of gas and stars in the $v_{\rm bc}=0\sigma_{\rm bc}$ are less apparent in the $v_{\rm bc}=2\sigma_{\rm bc}$ case. 
% even denser) stars and gas in $v_{\rm 
% bc}=2\sigma_{\rm bc}$. 
Although Fig.~\ref{Fig:box} shows just one example region, visual inspection of tens of highly star forming objects in the simulation boxes reveals a similar pattern. 
We quantify the extent of these effects on star formation below, as they have implications for the star formation rate estimation and the luminosity of these objects.

\subsection{Star formation and the Galaxy Stellar Mass Function}
\label{sec:GSMF}

\begin{figure}

\centering
   
\includegraphics[width=0.8\linewidth]{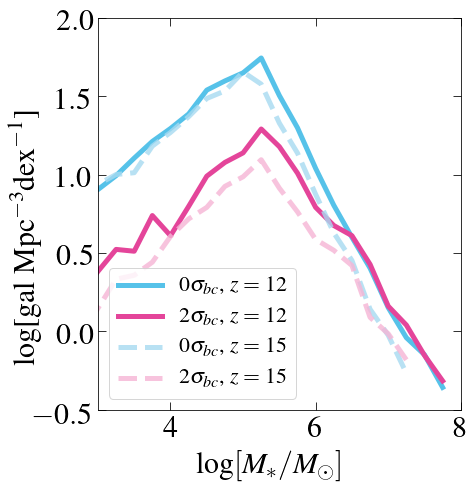}     
    \caption{Galaxy stellar mass function with (pink lines below) and without (blue lines above) stream velocity at $z=12$ (solid) and $z=15$ (dashed).  } \label{Fig:GSMF} 
\end{figure}
Given that the stream velocity is expected to delay and suppress 
small-scale
star formation at high redshift \citep[e.g.,][]{Schauer+23,Hegde+23,Conaboy+23}{}{}, and that \jwst can already provide constraints on the low-mass end of the GSMF at redshifts as high as $z\sim 8$ \citep[e.g.,][]{Navarro+23}{}{}, we calculate the GSMF in each simulation box to determine the expected contribution from the stream velocity.

\begin{figure}

\centering
   
\includegraphics[width=0.7\linewidth]{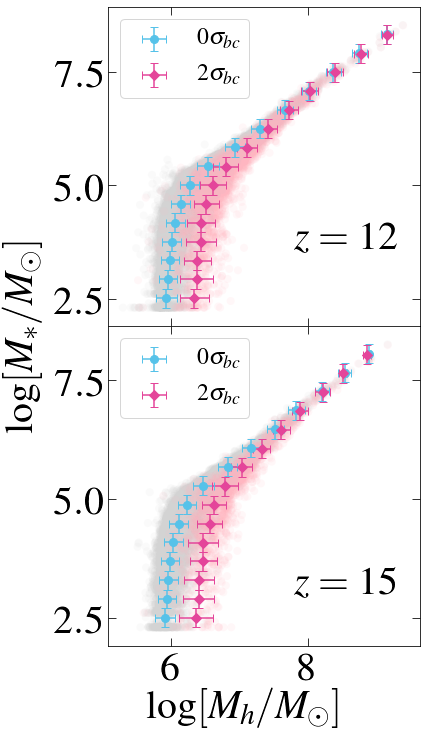}     
    \caption{{ Stellar mass ($M_*$) vs total halo mass ($M_h$) for the two runs at $z=12$ (top) and $z=15$ (bottom)}. 
    The data are binned along the y-axis and shown with error bars corresponding to one standard deviation. The $0\sigma_{bc}$ run is given in blue (left points), and the $2\sigma_{bc}$ run in pink (points to the right). Underneath the binned data, the actual distributions are shown, with the $0\sigma_{bc}$ runs shown in grey and the $2\sigma_{bc}$ runs shown in pink.   }\label{Fig:MsMh} 
\end{figure}

\begin{figure}

\centering
   
\includegraphics[width=0.8\linewidth]{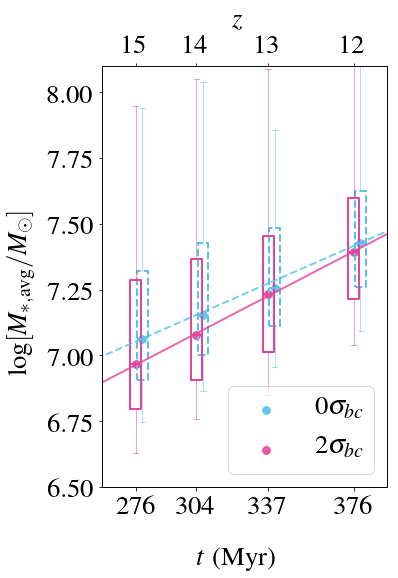}     
    \caption{ {Box plot of stellar mass in the most massive 100 galaxies in the simulation between $z=15$ and $z=12$. The medians are shown as circles and the boxes extend from the first to the third quartiles at each redshift. For clarity, the dot-dashed blue boxes corresponding to the no stream velocity case have been offset along the x-axis by $+3$ Myr.
Best fit lines to the medians  give,  $\log_{10}\left({M_*}/{M_\odot}\right)=0.0043 \log_{10}\left({t}/{\text{Myr}}\right) +5.78$ for the stream velocity run (solid line) and $\log_{10}\left({M_*}/{M_\odot}\right)=0.0036 \log_{10}\left({t}/{\text{Myr}}\right) +6.06$ for the no stream velocity run (dashed line). } This panel highlights the difference between the large masses shown 
    % in panel (a) 
    in Fig.~\ref{Fig:MsMh}
    as a function of redshift.  Whiskers extend to the most distant data point within 1.5 times the interquartile range from the box.} \label{Fig:SFR} 
\end{figure}

The results are presented in Fig.~\ref{Fig:GSMF}, where we show %plot 
the galaxy stellar mass function in our simulation volume for the runs with (pink lines) and without (blue lines) the stream velocity. The two redshifts examined ($z=12$ and $z=15$), are displayed with solid and dashed lines, respectively.
Above $M_*\gtrsim 10^{6.5}M_\odot$, the two curves agree. 
However, at lower masses, there are up to an order of magnitude more galaxies at each stellar mass in the run without the stream velocity. 

The suppression in the GSMF in the $2\sigma_{\rm bc}$ run can be attributed to two effects caused by the stream velocity. 
First, the number density of all halos is suppressed in regions of streaming, leading to a suppression of the GSMF \citep[see Fig.~\ref{Fig:STRatio} in App. \ref{ap:Clustering}, and, e.g.,][]{Naoz+12,Lake+21}.
Additionally, we show in
Fig.~\ref{Fig:MsMh} 
the stellar mass as a function of halo mass for objects in each run. 
For a given stellar mass, the stars lie in a much larger halo in regions of streaming than without streaming. 
Thus, larger halos, which are less abundant according to standard halo mass function, host galaxies of the same stellar mass for $M_*\lsim10^{6.5}M_\odot$, contributing fewer galaxies to the GSMF. 
As expected for a study with no feedback effects included, our simulated $M_*/M_h$ at most masses is ten times or more higher than is found by studies that include radiative and supernova feedback effects at similar redshift \citep[e.g.,][]{Xu+16b, Yeh+23}{}{}.
Typically, when feedback is included, $M_*/M_h\sim10^{-3}$ \citep[e.g.,][]{Garaldi+22,Yajima+23}{}{}, while our study has objects mostly with $10^{-2}\lsim M_*/M_h\lsim10^{-1}$.

As demonstrated by Fig.~\ref{Fig:MsMh}, while low stellar mass objects in the case with stream velocity have higher DM mass, the stream velocity and classical cases converge at large masses. 
Furthermore, the mass at which the two converge decreases with decreasing redshift. 
In other words, the blue ($0\sigma_{\rm vbc}$), and the red ($2\sigma_{\rm vbc}$) points, lie on top of each other for $M_*\geq 10^{7.0}M_\odot$ at $z=15$.
However, at $z=12$, they converge at $M_*\geq 10^{7.25}M_\odot$, 
implying a fast SFR for the high masses with stream velocity case, between these two snapshots at $z=15$ and $z=12$. 

This behavior is further illustrated in Fig.~\ref{Fig:SFR}, where the time evolution of the average stellar mass is shown for the most massive galaxies by stellar mass. 
As depicted, the star formation rate (the slope of the Figure) in the stream velocity case is larger.  
Thus, although the star formation is suppressed for most galaxies in the stream velocity simulation run until $z\sim 15$, the star formation {\it rate} at large masses with a large stream velocity is heightened between $z=15$ and $z=12$. 
 Specifically, at $z=12$, the mean star formation rate among galaxies with $M_*>10^7 M_\odot$ that actively formed stars between $z=13$ and $z=12$ was around twice as high in the stream velocity run ($0.986 M_\odot$ yr$^{-1}$) as the run without the stream velocity ($0.453 M_\odot$ yr$^{-1}$).

\subsection{The UV luminosity function}
\label{sec:uvluminosity}

\begin{figure*}

\centering
   
\includegraphics[trim={0 0 0 0},width=0.75\linewidth]{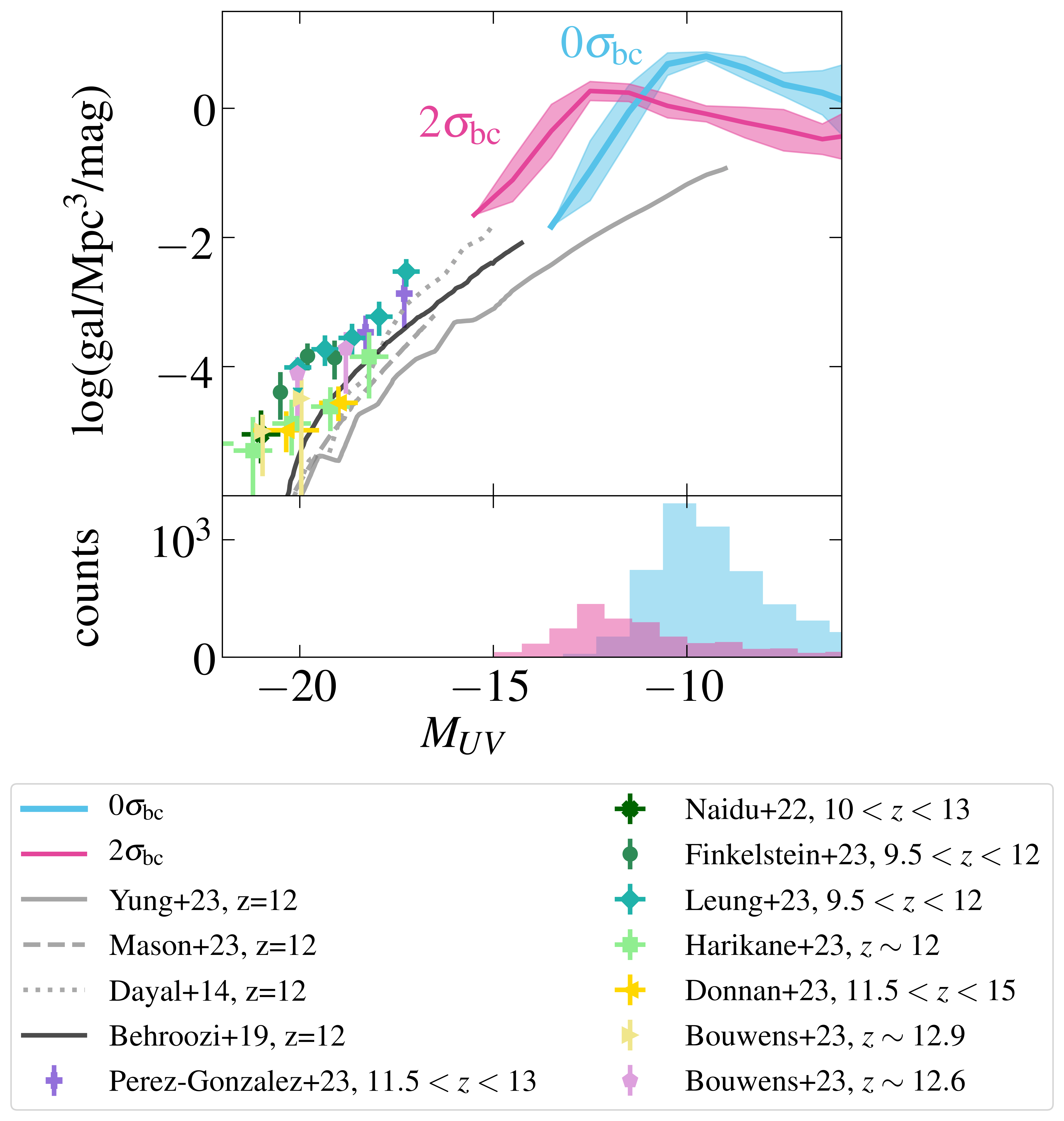}     
    \caption{{\bf Top:} UV Luminosity Function at $z=12$ with (pink) and without the stream velocity (blue) compared to the theoretical models of \cite{Yung+23}, \cite{Mason+23}, \cite{Dayal+14}, and observational constraints close to the redshift presented from \cite{PerezGonzalez+23,Naidu+22candidates,Finkelstein+23ceersI,Leung+23uvlf,Harikane+23uvlf,Donnan+23uvlf,Bouwens+23uvlf,Bouwens+23uvlfevolution}. 
    The UVLF is then averaged over bins with $\Delta M_{UV}=1$, with the shaded regions showing the minimum and maximum value in that box. {\bf Bottom: }Distributions of galaxies by AB absolute UV magnitude with and without the stream velocity between $z=13$ and $z=12$.
    }\label{Fig:UVLF} 
\end{figure*}

Given the above prediction that the star formation rate is higher in patches of the Universe with high stream velocity at the redshifts probed by our simulation, we expect that high redshift dwarfs will be brighter in UV emission in those patches. 

Thus, as a second probe of the population of dwarf galaxies resulting from a streaming scenario versus the classical scenario, %standard, 
we calculate the rest frame absolute UV magnitude $M_{UV}$ of our objects at $z=12$. This is done based on the SFR according to the procedure described in \S~\ref{sec:Methods}. 

In the bottom panel of Fig.~\ref{Fig:UVLF} we show the distribution of objects in our simulation box binned by $M_{UV}$.
For objects fainter than $M_{UV}\sim -12$, the UV magnitudes follow the same trend as the stellar masses, with the stream velocity function only $1/6$ of the no streaming case. 
However, a turnover occurs above $M_{UV}\sim -12$, and the stream velocity counts are enhanced by a factor of over an order of magnitude at $M_{UV}\sim -14$. 

We show the UVLF in regions with and without streaming in the top panel of Fig.~\ref{Fig:UVLF}.  
The trend in the distributions is reflected in the UVLF at the bright end of our sample as a turnover where the UVLF in the $2\sigma_{bc}$ region is greater than the no streaming function.  
The enhancement in the UVLF below $M_{UV}\sim -12$ in the presence of the stream velocity can be counterintuitive given the suppression in star formation and number density caused by the stream velocity in most other contexts, including the GSMF. 
As evident in Fig.~\ref{Fig:MsMh}, there is no underlying overabundance of more massive galaxies by stellar mass (which have higher SFR) due to the stream velocity--at these larger masses, there are roughly equal abundances in both simulation runs. 
Instead, several effects play a role in boosting the UV luminosity of these more massive galaxies. 
Importantly, the UV luminosity in our model is proportional to the SFR of galaxies via Eq.~(\ref{eq:Luv}). 
Here, we discuss how the stream velocity in this model leads to a boost in the star formation rate of  dwarf galaxies with slightly higher stellar mass ($M_*\gsim 10^{6.5}M_\odot$).

Firstly, in regions with highly supersonic streaming flow, the small-scale structure of gaseous clumps is washed away \citep[as shown by e.g.,][]{Tes+10a, Fialkov+11,BD,Maio+11, OLMc12,Naoz+11a,Tes+10b,TanakaLi+13,Tanaka+14}{}{}. 
An example of this suppression is evident in the small region of our simulation snapshot at $z=12$ depicted in Fig.~\ref{Fig:box}, where at least thirteen small gaseous clumps are present around the central galaxy in the $0~\sigma_{bc}$ run while only five similar clumps are present in the $2\sigma$ run. 
Thus, star formation without the stream velocity can proceed in small minihalos of condensed gas, contributing to the behavior of the low mass GSMF in Fig.~\ref{Fig:GSMF}.
Small clumps of star formation corresponding to these gaseous clumps are seen in the right-hand panel of Fig.~\ref{Fig:box}.
In regions with the stream velocity, however, the same overall cosmic density of gas is still present. 
Having been swept away from clumps that would allow it to form small clusters, the diffuse gas streams back onto larger halos thanks to their gravitational influence over the region as the stream velocity decays. 
Thus, star formation is concentrated in larger dwarf galaxies rather than smaller minihalos. 
This suppresses the extreme faint end of the UVLF while increasing the luminosity of the larger dwarfs, producing the trend in Fig.~\ref{Fig:UVLF}.

Furthermore, previous studies such as \cite{Stacy+10,Greif+11, Schauer+17, Schauer+23} found a delay in star formation in regions of streaming. 
\cite{Schauer+23} found that by $z=8$, dwarf galaxies in their simulations with the stream velocity had reached roughly the same stellar mass as counterparts in regions without streaming. 
For example, Figure 4 of that study depicts a dwarf galaxy that reaches $\sim 10^7M_\odot$ by the end of the simulation and grows at a rate of $\sim 0.031 M_\odot$ yr$^{-1}$ between $z=15$ and $z=12$, while during the same period in the run with the stream velocity it grows at $\sim 0.038 M_\odot$ yr$^{-1}$. 
The authors note a modest effect of higher star formation rates in their streaming run overall prior to $z\approx 8$, at which point the difference in mass between the runs becomes mostly stochastic. 
If stream velocity-induced dwarfs indeed catch up to the stellar mass of the non-streaming population by the epoch of reionization after a delay in the onset of star formation, then they undergo a period of faster SFR than the non-streaming case.

The UVLF at $z=12$, which roughly tracks the instantaneous SFR at the redshift of observation, would be catching these objects in a period where they experience enhanced star formation thanks to the mechanism of concentrating gas into larger halos described above. 
 We note that the baryonic feedback resulting from star formation episodes in these structures is not investigated here. 
 Our UVLF can be understood as an upper limit in both cases given the likely reduction in star formation efficiency that would result from the inclusion of feedback in the simulation. 
A heightened SFR may lead to increased disruption due to feedback, however, as in Fig~\ref{Fig:MsMh}, dwarf galaxies with the stream velocity sit in larger DM halos in streaming cases which may prevent disruption.
The stream velocity may thus play a role in the ``burstiness" of star formation at early times.

We overplot our results to several other theoretical models in the literature from \cite{Dayal+14,Behroozi+19,Yung+23,Mason+23} at $z=12$. This Figure, thus, demonstrates that the high-resolution investigation presented here allows us to reach the extremely faint end of the luminosity function. 
 As this study does not include feedback effects, it is also relevant to note that our predicted $0\sigma_{\rm bc}$ curves lie above the simulated UVLFs of radiation hydrodynamical simulations in the literature that incorporate supernova and other feedback effects at similar redshift \citep[e.g.,][]{Xu+16b,Borrow+23}{}{}, typically by around an order of magnitude. 
We note that none of these studies included the stream velocity effect.

In addition, we overplot some current observational \jwst ~constraints. 
Our box size does not include enough objects bright enough to fall within the observable range of current \jwst  ~surveys. 
However, JWST studies focused on the faint-end of the UVLF, such as \cite{PerezGonzalez+23} and \cite{Leung+23uvlf} are within a few magnitudes of our brightest galaxies. 
Given a strong lensing scenario, where magnifications allow for observations of objects that are an order of magnitude fainter than these studies, a measurement of the UVLF that reaches the predictions here may soon be possible.

\section{Discussion and Conclusion}
\label{sec:Discussion}
Our results imply that JWST observations of the faint end of the UVLF are close to a regime sensitive to the stream velocity. 
\textit{In particular, we predict that a high-stream velocity patch will contribute to a higher UVLF compared to the UVLF predicted by classic structure formation.}
This work also confirms that a strong suppression of galaxies with low stellar mass ($M_*\lsim 10^{6.5}\, M_\odot$) is present in the early Universe  due to the stream velocity. 

Although this faint regime has not yet been detected, given that dwarf galaxies may contribute significantly to the production of ionizing photons at high redshift, the stream velocity effect may contribute to a spatial modulation in the progression of the early phases of reionization. 
Additionally, the suppression may persist to lower redshifts beyond the end of our simulation, where it could be detected by JWST or future missions that survey dwarf galaxies. 

 Our simulations do not include any baryonic feedback effects, and as expected this increases our star formation efficiency in both runs by an order of magnitude in comparison to studies that include feedback effects. For example, both our stellar/halo mass function and our UVLF lie at least ten times above the simulated halos of \citet[][]{Xu+16b}{}{}, which include metal-free and metal-enriched star formation and feedback. Studies that investigated the stream velocity in the presence of feedback \citep[e.g.,][]{Schauer+23}{}{} did not include a different prescription for feedback between the two cases and did not report a difference in the effects of feedback with and without the stream velocity. We caution that because feedback has some degenerate effects with a high stream velocity value, a boost or suppression in SFR of a single object is not enough to distinguish the stream velocity scenario. Instead, it will be necessary to investigate the overall population of objects, especially over large spatial scales.
Given that the stream velocity coherence scale is on the order of a few tens of Mpc, observational measurements of the UVLF drawn from cosmic volumes much larger than the coherence scale will represent objects drawn from both low and high values of the stream velocity. 
Thus, the individual curves shown in our Fig. \ref{Fig:UVLF} will not be resolved except by observations below the coherence scale. 
When comparing surveys selected from volumes larger than the coherence scale, the stream velocity effect should contribute to an overabundance of UV bright galaxies, scattering points above the standard $\Lambda$CDM curve.
Our results predict that observations from regions comparable and smaller than the coherent length at a certain redshift would fall along a curve corresponding to a single value of the stream velocity.
The stream velocity varies on  spatial scales $>$ few Mpc and varies randomly in magnitude and direction in space following a Maxwell-Boltzmann distribution \citep[e.g.,][]{Fialkov14}{}{}. 
 We note that this study only investigates the effects with a single value of the stream velocity as a proof-of-concept, and we leave detailed studies of intermediate or larger values to future work.

We suggest that as data sets at high redshift grow in completeness with future observations, it will be possible to undertake a statistical comparison of a larger cosmological volume to determine the presence of the stream velocity. Furthermore, since high-density peaks often correlate to large stream velocity values \citep[e.g.,][]{Tes+10b}, we expect that deep field JWST observations will likely probe volumes formed in the presence of the stream velocity.
For an ultra deep field survey such as NGDEEP \citep[e.g.,][]{Bagley+23}{}{} observing a highly magnified field ($\mu\sim10-50$) that reaches $M_{UV}\sim -14$, we expect an enhanced luminosity function at the faint end.

JWST has opened a window into the high redshift Universe and the earliest stages of galaxy formation, allowing for cosmological tests at earlier times than previously possible. 
While current interest often focuses on the observed discrepancies at the bright end of the UVLF, we propose that the faint end of the UVLF contains information about baryonic physics and the interaction with dark matter at early times. 
The stream velocity is a robust prediction of $\Lambda$CDM and is not present or decays rapidly in some alternative DM models \citep[e.g.,][]{Nadler+19,Maamari+21,Driskell+22}{}{}, thus, our methods represent a new test of $\Lambda$CDM structure formation and may help constrain alternative models. 
Such observed faint-end fluctuations in the UVLF on the coherence scale of the stream velocity would constitute the first high-redshift evidence of the presence of the stream velocity in the early Universe.

\section*{Acknowledgements}
The authors would like to thank Steve Furlanetto and Sahil Hegde for constructive conversations and feedback. 
 C.E.W.  acknowledges the support of the National Science Foundation Graduate Research Fellowship and the University of California, Los Angeles. C.E.W., W.L., S.N., Y.S.C, B.B., F.M., and M.V. thank the support of NASA grant No. 80NSSC20K0500 and the XSEDE/ACCESS AST180056 allocation, as well as, the UCLA cluster Hoffman2 for computational resources.
This material is based upon work supported by the National Science Foundation Graduate Research Fellowship Program under Grant No. DGE-2034835. Any opinions, findings, and conclusions or recommendations expressed in this material are those of the author(s) and do not necessarily reflect the views of the National Science Foundation.

\software{The authors would like to acknowledge the Astropy \citep{astropy:2013, astropy:2018, astropy:2022}, Web Plot Digitizer \citep[][]{Rohatgi2022}, {\tt yt} \citep[][]{Turk+11}{}{}, NetworkX \citep[][]{NetworkX}{}{}, and Matplotlib \citep[][]{Matplotlib}{}{} software packages, which were used in this work.}

\bibliography{cosmo}{}
\bibliographystyle{aasjournal}
\begin{appendix}

\section{Clustering correction factor}
\label{ap:Clustering}
\begin{figure}

\centering
   
\includegraphics[width=0.95\linewidth]{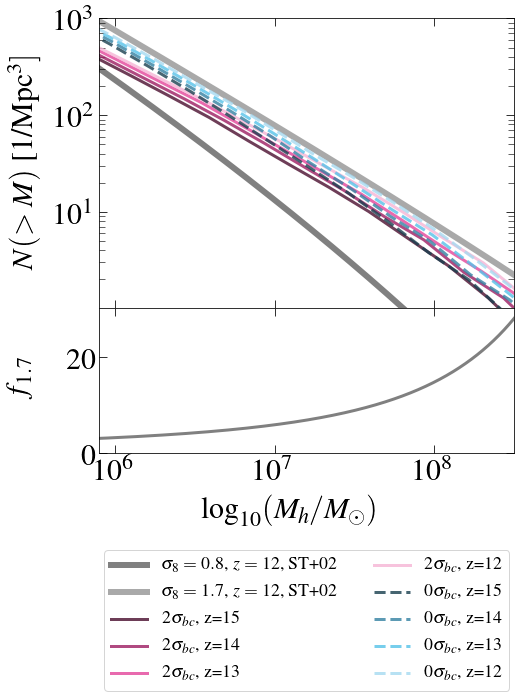}     
    \caption{{\bf Top Panel:} The number of halos greater than mass {\bf $M$ } for the standard value $\sigma_8 = 0.8$ (dark grey) and $\sigma_8=1.7$ (light grey) used in the simulations presented here.
    The halo mass function from \cite{ST+02} is used. Also shown are the numerical results from our simulations at a selection of redshifts  (pink and blue dashed lines). {\bf Bottom Panel:} The ratio between $n(>M)$ at $\sigma_8 = 0.8$  and $\sigma_8=1.7$ ($f_{1.7}$) as a function of halo mass. This ratio corresponds to the correction factor used to calibrate our UVLF. }\label{Fig:STRatio} 
\end{figure}
In our simulations, the value of $\sigma_8$ is increased from the nominal value of $\sigma_8=0.8$ to $\sigma_8=1.7$ in order to increase the statistical power of our study without affecting the underlying cosmology. 
However, this results in an overabundance of halos. 
In order to describe the number density of objects in a typical region with $\sigma_8=0.8$, we 
apply an analytically calculated correction based on the \cite{ST+02} halo mass function. 
Following \cite{NB07}, we calculate the cumulative number density of halos  
\begin{equation}
    n(>M_{min})=\int_0^\infty \frac{\text{d}n}{\text{d}M}\text{d}M,
\end{equation}
using the \cite{ST+02} halo mass function with $\sigma_8=1.7$ and $\sigma_8$ at $z=12$:
\begin{equation}
    \frac{\text{d}n}{\text{d}M}=\frac{\rho_0}{M} f_{ST}
    \left|{\frac{\text{d} S}{\text{d}M}}\right|,
\end{equation}
where 
\begin{equation}
\label{eq:STfrac}
    f_{ST}(\delta_c,S) = A' \frac{\nu}{S} \sqrt{\frac{a'}{2\pi}}\left[1+\frac{1}{(a'\nu^2)^{q'}}\right]\exp{\left(\frac{-a'\nu^2}{2}\right)}.
\end{equation}
Here, $\nu=\delta_c/\sqrt{S}$, $a'=0.75,$ $q'=0.3,$ $A'=0.322$ and the critical collapse overdensity ($\delta_c(z)$) and the variance calculated from the power spectrum ($\sigma^2(M,z)$) are the arguments of Eq.~\ref{eq:STfrac} \citep[][following \citet{ST+02}]{NB07}{}. 

Our simulation number counts with $v_{bc}=0\sigma_{vbc}$ are consistent with the Sheth \& Tormen value  (see top panel of Fig.~\ref{Fig:STRatio} comparing the numerical number counts to the analytical prediction). 
The correction factor is a function of mass. 
To correct the UVLF, we calculate the average mass in each bin and divide by the correction factor corresponding to the average mass. 
This introduces an uncertainty in our predicted value due to the range of masses in each magnitude bin. 
The correction factor is shown in Fig.~\ref{Fig:STRatio}
in the bottom panel.

\section{Example buildup of stellar mass}
\label{ap.stellarmass}
\begin{figure}

\centering
   
\includegraphics[width=0.8\linewidth]{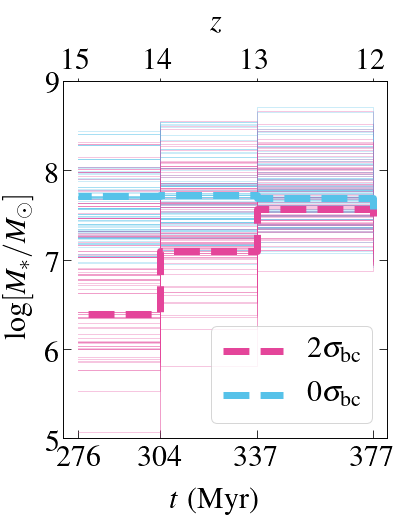}     
    \caption{
    Stellar mass of massive objects between $z=12$ and $z=15$ with the stream velocity (pink) and the same object without the stream velocity (blue). An example object (the same object shown in Fig.~\ref{Fig:box} is shown in the thick dashed lines.
    In the simulation run with stream velocity (shown in pink), the object grows from $10^{6.39}M_\odot$ to $10^{7.49}M_\odot$, whereas without the stream velocity (blue), the object has already formed the bulk of its stellar mass by that time. 
    }\label{Fig:Mstarobj} 
\end{figure}

Figure~\ref{Fig:Mstarobj} shows the stellar mass as a function of redshift for an example object in the simulation (solid dashed lines).
In a period between $z=15$ and $z=12$, the object in the run with the stream velocity builds up the majority of its stellar mass, whereas its counterpart in the no stream velocity run does not undergo significant star formation because it already accrued the bulk of its stellar mass prior to that time.  
By $z=12$, the object with the stream velocity has reached a similar mass to the object in the $0\sigma_{\rm bc}$ run. In the background, similar tracks comparing objects with and without the stream velocity are shown in faint solid lines.
\end{appendix}
\end{document}